\documentclass{aamas2017}
\usepackage[protrusion=true,spacing=true]{microtype}
\usepackage[group-separator={,}]{siunitx}
\usepackage{amsmath,amsfonts}
\usepackage{graphicx,subfig}
\usepackage{cite,setspace}
\usepackage[usenames]{xcolor}
\usepackage[noend]{algorithmic}
\usepackage{algorithm}
\usepackage{enumitem}

\usepackage[absolute,overlay]{textpos}
\newcommand{\typ}{\theta}
\newcommand{\Typ}{\Theta}
\newcommand{\Up}{\Phi^t}
\newcommand{\bayesup}{P(\typ_j | H_i^t) \propto P(a_j^{t-1} | H_i^{t-1}, \typ_j, p^t) P(\typ_j | H_i^{t-1})}
\newcommand{\Mem}{\textit{Mem}}
\newcommand{\Loc}{\textit{Loc}}
\newcommand{\Des}{\textit{Dest}}
\newcommand{\Tar}{\textit{Targ}}
\newcommand{\com}[1]{{\color{gray}\it // #1}}
\newcommand{\cor}{\textsc{Cor}}
\newcommand{\rnd}{\textsc{Rnd}}

\begin{document}

	\title{Reasoning about Hypothetical Agent Behaviours \\ and their Parameters \vspace{-5pt}}

	\numberofauthors{2}

	\author{
		\alignauthor
		Stefano V. Albrecht \\
		\affaddr{Department of Computer Science} \\
		\affaddr{The University of Texas at Austin} \\
		\affaddr{Austin, TX 78712, USA} \\
		\email{svalb@cs.utexas.edu}
		\alignauthor
		Peter Stone \\
		\affaddr{Department of Computer Science} \\
		\affaddr{The University of Texas at Austin} \\
		\affaddr{Austin, TX 78712, USA} \\
		\email{pstone@cs.utexas.edu}}

	\maketitle

	\begin{abstract}
Agents can achieve effective interaction with previously unknown other agents by maintaining beliefs over a set of hypothetical behaviours, or \emph{types}, that these agents may have. A current limitation in this method is that it does not recognise \emph{parameters} within type specifications, because types are viewed as blackbox mappings from interaction histories to probability distributions over actions. In this work, we propose a general method which allows an agent to reason about both the relative likelihood of types \emph{and} the values of any bounded continuous parameters within types. The method maintains individual parameter estimates for each type and selectively updates the estimates for some types after each observation. We propose different methods for the selection of types and the estimation of parameter values. The proposed methods are evaluated in detailed experiments, showing that updating the parameter estimates of a single type after each observation can be sufficient to achieve good performance.
	\end{abstract}





	\section{Introduction} \label{sec:intro}

An important open problem in multi-agent systems is the design of autonomous agents that can quickly and effectively interact with other agents when there is no opportunity for prior coordination, such as shared world models and communication protocols \cite{als2016jaamas,skkr2010,bm2005}. Several works addressed this problem by proposing methods which utilise beliefs over a set of hypothetical behaviours for the other agents\cite{acr2016aij,ar2014,bs2015,bsk2011,cdzc2014,sbl2005}. Behaviours in this approach are specified as \emph{types}, which are blackbox mappings from interaction histories to probability distributions over actions. If the types are sufficiently representative of the true behaviours of other agents, then this method can lead to rapid adaptation and effective interaction in the absence of explicit prior coordination \cite{bs2015,ar2013}.

There is, however, a current limitation in this type-based method, which is that it does not recognise \emph{parameters} within types. Complex behaviours often involve various continuous parameters which govern certain aspects of the behaviour. For example, reinforcement learning methods often use learning, discounting, and exploration rates \cite{sb1998}. If we were to use such a method as a type, we would have to instantiate its parameters to some fixed values. Thus, an agent that wants to account for $n$ different parameter settings will have to reason about $n$ instances of the same type whose only difference is in their parameter values. This, however, is very inefficient as it leads to redundancy in space (storing $n$ copies of the type) and time (computing the outputs of $n$ copies).

Our goal in this work is to devise a method which allows an agent to reason about both the relative likelihood of types \emph{and} the values of their parameters. To be useful in practice, this reasoning should be efficient and allow for any bounded continuous parameters, without a need for the user to specify maximum likelihood estimators for the individual parameters.

We show that the problem of space redundancy is typically unavoidable because the internal state of a type may depend on both the history of observations \emph{and} the parameter values. Regarding the time requirements, due to the blackbox nature of types, the only way to ascertain the effect of a specific parameter setting is to evaluate the type with that parameter setting. Thus, our goal is to minimise the number of type evaluations while achieving a useful and robust estimate of the type's true parameter setting. We propose a general method which maintains individual parameter estimates for each type and \emph{selectively updates} the estimates for some types after each observation. We propose different methods for the selection of types and the estimation of parameter values. The proposed methods are evaluated in the level-based foraging domain \cite{ar2013}, where they achieved substantial improvements in task completion rates compared to random estimates, while updating only a single parameter estimate in each time step.

	\section{Model \& Objective} \label{sec:prel}

We consider an interaction process with two or more agents. The process starts at time $t = 0$. At time $t$, each agent $i$ receives a signal $s_i^t$ and independently chooses an action $a_i^t$ from some countable set of actions $A_i$. The signal $s_i^t$ may encode information about the state of the environment, a private reward, etc. We leave the precise structure and dynamics of $s_i^t$ open. This process continues indefinitely or until some termination criterion is satisfied.

The probability with which action $a_i^t$ is chosen is given by $P(a_i^t | H_i^t, \typ_i, p)$, where $H_i^t = (s_i^0,...,s_i^t)$ is  agent $i$'s history of observations, $\typ_i$ is $i$'s \emph{type}, and $p = (p_1,...,p_n)$ is a vector of \emph{continuous parameters} in $\typ_j$. Each parameter $p_k$ takes a fixed value from some bounded interval $[p_k^{\min},p_k^{\max}] \subset \mathbb{R}$. To simplify the exposition, we assume that all types have the same number of parameters, but in general this need not be the case. Which type $\typ_j$ a parameter vector $p$ belongs to is disambiguated from context.

We control a single agent, $i$, which reasons about the behaviour of another agent, $j$. We assume that $i$ knows $j$'s action space $A_j$ and that it can observe $j$'s past actions, i.e. $a_j^{t-1} \in s_i^t$ for $t > 0$. The true type of $j$, denoted $\typ_j^*$, and its true parameter values, $p^*$, are unknown to $i$. However, $i$ has access to a finite set of \emph{hypothetical types} $\typ_j \in \Typ_j$, with $\typ_j^* \in \Typ_j$. We furthermore assume that $i$ has all information relevant to $j$'s decision making, so that $H_j^t$ is a function of $H_i^t$ and we can write $P(a_j^t | H_i^t, \typ_j, p)$.

The goal in this work is to devise a method which allows agent $i$ to reason about the relative likelihood of types $\typ_j \in \Typ_j$ \emph{and} the values of their parameters $p$, based only on agent $j$'s observed actions.


	\section{Markovian Parameters} \label{sec:markov}

Types are often implemented as Markov chains, such that the choice of action depends only on the current signal $s_j^t$ and a current internal state $w_j^t$ of the type, i.e. $P(a_j^t | s_j^t, w_j^t, \typ_j, p)$. The information contained in $s_j^t$ is then incorporated into the next state $w_j^{t+1}$, usually by aggregating the information within a collection of variables inside the state.

For types which are realised in this way, it is important to note that the internal state of the type may depend on both the history of observations \emph{and} the parameter values. To illustrate this, consider a simple Q-learning agent \cite{wd1992} which uses three parameters, $\alpha, \gamma, \epsilon \in [0,1]$. Its internal state is defined by a matrix, $Q$, which is used to compute and store expected payoffs for state-action pairs. This matrix is updated at each time step as
\begin{equation*}
	Q(s,a) \gets (1-\alpha) Q(s,a) + \alpha \left[ r + \gamma \max_{a'} Q(s',a') \right]
\end{equation*}
where $s,a$ is the previous state-action pair, $r$ is some reward, and $s'$ is the new state. Given a state $s$, the agent chooses an action in $\arg\max_a Q(s,a)$ with probability $1-\epsilon$, and a random action otherwise. In this example, the values of $Q$ depend on the history of observed states and rewards \emph{and} the values of $\alpha, \gamma$.

This dependence on parameter values has important consequences for space requirements. Suppose we use the Q-learning agent as a type $\typ_j$ and fix its parameter setting to some values $p$. Its internal state $w_j^t$, defined by $Q$, will depend on past observations and $p$. Now, if we change the parameter setting to $p' \neq p$ at some time $t$, we have a potential inconsistency in that $P(a_j^t | s_j^t, w_j^t, \typ_j, p')$ may not be equal to $P(a_j^t | H_j^t, \typ_j, p')$, since $w_j^t$ has thus far been updated using $p$. Therefore, to ensure correct probabilities, we may have to adjust $w_j^t$ to conform to the new parameter setting $p'$. In general, this can be done by recomputing the internal state ``from the ground up'' using the new parameter setting. However, more efficient methods may be possible depending on how the internal states are influenced by parameters.

We adopt the naming convention and say that parameters $p$ of type $\typ_j$ are \emph{Markovian} if $\typ_j$'s action probabilities are independent of past values of $p$ given their current values, i.e.
\vspace{-1em}
\begin{equation} \label{eq:markov}
	P(a_j | H_j^t, \typ_j, p^t,p^{t-1},...,p^0) = P(a_j | H_j^t, \typ_j, p^t)
\end{equation}
where $p^{\tau}$ are the parameter values at time $\tau$. Hence, the parameters in the Q-learning example (specifically $\alpha,\gamma$) are not Markovian since they directly influence the values of $Q$.

	\section{Learning Parameters in Types} \label{sec:algo}

We propose a method whereby agent $i$ maintains individual parameter estimates for each hypothetical type $\typ_j \in \Typ_j$ and selectively updates the estimates after each observation.

The method starts with an initial belief $P(\typ_j | H_i^0)$ which specifies the relative likelihood (probability) that agent $j$ has type $\typ_j$. In addition, for each type $\typ_j \in \Typ_j$, it maintains an initial parameter estimate $p^0$ within the respective value bounds. Then, at each time $t > 0$, the method selects a subset of types $\Up \subset \Typ_j$ and obtains a new parameter estimate $p^t$ for each $\typ_j \in \Up$. (Sections~\ref{sec:select} and \ref{sec:estim} propose methods for each of these operations.) If the parameters of a type $\typ_j \in \Up$ are non-Markovian, then the internal state of $\typ_j$ may have to be adjusted to conform to the new parameter estimate (cf. Section~\ref{sec:markov}). The parameter estimates of types not in $\Up$ remain unchanged. Given the estimate $p^t$ for type $\typ_j$, the current belief is updated via
\begin{equation} \label{eq:bayesup}
	\bayesup
\end{equation}
and the method continues in this fashion (cf. Algorithm~\ref{alg:main}).

\begin{algorithm}[t]
	\textbf{Given:} type space $\Typ_j$, initial belief $P(\typ_j | H_i^0)$ and parameter estimate $p^0$ for each type $\typ_j \in \Typ_j$ \\[3pt]
	\textbf{Repeat} for each $t > 0$:
	\begin{algorithmic}[1]
		\algsetup{linenodelimiter=:\ }
		\setstretch{1.2}
		\STATE Select a subset $\Up \subset \Typ_j$ for parameter updates
		\STATE For each $\typ_j \in \Up$:
		\STATE \quad Obtain new parameter estimate $p^t$ for $\typ_j$
		\STATE \quad If $p^t$ non-Markovian, adjust internal state of $\typ_j$
		\STATE Set $p^t = p^{t-1}$ for all $\typ_j \not\in \Up$
		\STATE For each $\typ_j \in \Typ_j$, update belief:
		\vspace{-3pt}
		\begin{displaymath}
			\bayesup\vspace{-3pt}
		\end{displaymath}
	\end{algorithmic}
	\caption{Selective parameter estimation in types}
	\label{alg:main}
\end{algorithm}

The use of point estimates of parameters effectively allows us to use Algorithm~\ref{alg:main} as a pre-routine on top of an existing implementation $A$ of the type-based method (e.g. \cite{bskr2013,ar2013}). That is, at each time $t > 0$, we first execute Lines 1-5 to set the parameter values for each type, after which Line 6 executes $A$ to update the belief and perform the planning step. From the perspective of $A$, there is formally no difference in the types since their parameters were set externally.

However, using point estimates can also cause a potential problem in our setting: it may generally be the case that $P(a_j^{t-1} | H_i^{t-1}, \typ_j^*, p^t) = 0$ while $P(a_j^{t-1} | H_i^{t-1}, \typ_j^*, p^*) > 0$.\footnote{\small As an example, consider the Q-learning agent from Section~\ref{sec:markov} and set $\epsilon^t = 0$, $\epsilon^* = .5$, and $a_j^{t-1} \not\in \arg\max_a Q(s,a)$.} The latter can cause $P(\typ_j^* | H_i^t)$ to prematurely converge to zero, even though we may learn the correct parameter values $p^*$ at a later time. To prevent this, we assume that for any $\typ_j \in \Typ_j$, if $P(a_j^{t-1} | H_i^{t-1}, \typ_j, p)$ is positive for some $p$, then it is positive for all $p$. In practice, this can be ensured by using close-to-zero probabilities instead of zero probabilities.

		\subsection{Selecting Types for Parameter Updates} \label{sec:select}

Since we do not know which type in $\Typ_j$ is the true type $\typ_j^*$, the safe choice of $\Up$ is to update the parameter estimates of all types in $\Typ_j$. However, this is also the most costly choice in terms of computation time. On the other hand, we may minimise computation costs by updating parameter estimates only for some subset $\Up \subset \Typ_j$, but this carries the risk that $\typ_j^*$ may not be included in $\Up$. In this sense, we view the choice of $\Up$ as a decision problem which balances exploitation (i.e. choosing types which are in some sense expected to benefit the most from an update) and exploration. We propose two approaches to make this choice, which entail different notions of exploitation, exploration, and risk.

			\subsubsection{Posterior Selection} \label{sec:select-post}

The first approach is to select types which are believed to be most likely, with the expectation that one of them is the true type. Here, exploitation amounts to choosing types $\typ_j \in \Typ_j$ which have maximum probability $P(\typ_j | H_i^{t-1})$. However, depending on the observation history $H_i^{t-1}$ and parameter estimates $p^0,...,p^{t-1}$, there is a risk that $P(\typ_j | H_i^{t-1})$ assigns high probability to incorrect types $\typ_j \neq \typ_j^*$. This can lead to premature convergence of beliefs to incorrect types if we do not update the parameter estimates of the true type $\typ_j^*$. Thus, exploration in this approach means choosing types which currently seem less likely than other types. To balance exploitation and exploration, we propose to sample $\Up$ from the belief $P(\typ_j | H_i^{t-1})$.

			\subsubsection{Bandit Selection} \label{sec:select-bandit}

The second approach is to select types according to their expected change in parameter estimates after the new observation is accounted for. This is predicated on the assumption that parameter estimates will converge, so that exploitation entails selecting types which are expected to make the largest leaps toward convergence. The risk is that the parameter estimates for some types, including the true type $\typ_j^*$, may not change significantly until certain observations are made. Hence, exploration entails choosing types even if their parameter estimates are not expected to change much.

To balance exploitation and exploration, we can frame this approach as a multi-armed bandit problem \cite{r1952}. In the general setting, there are $k$ arms to choose from at each time step $t$, and each choice results in a reward $r^t$ drawn from an unknown distribution associated with the chosen arm. The goal is to choose arms so as to maximise the sum of rewards. In our case, the arms represent the types in $\Typ_j$ and we define the reward $r^t$ after updating the parameter estimate of type $\typ_j$ as the normalised L1 norm
\begin{equation} \label{eq:bandit}
	r^t = \eta^{-1} \sum_{k=1}^n | p_k^t - p_k^{t-1} |, \hspace{10pt} \eta = \sum_{k=1}^n p_k^{\max} - p_k^{\min}.
\end{equation}
Thus, rewards are in the range $[0,1]$, where a reward of 0 means no change in the parameter estimate and a reward of 1 represents maximum change. Several algorithms exist which solve this problem, subject to different assumptions regarding the distribution of rewards (e.g. \cite{acf2002,kmrv1998}). In our case, the reward distributions of arms are independent but possibly changing over time (e.g. if estimates converge). Therefore, one should also consider algorithms designed for changing reward distributions (e.g. \cite{fm2004,acfs1995}).


		\subsection{Estimating Parameter Values} \label{sec:estim}

We propose three different methods for the estimation of parameter values $p^t$ of a type $\typ_j$. For notational convenience, we define $f(p) \doteq P(a_j^{t-1} | H_i^{t-1}, \typ_j, p)$.

		\subsubsection{Approximate Gradient Ascent} \label{sec:estim-aga}

The idea in this method is to update parameter estimates by following the gradient of a type's action probabilities with respect to the parameter values. Formally, the estimate is updated as $p^t = p^{t-1} + \lambda^t \,\nabla f(p^{t-1})$, where $\nabla f$ denotes the gradient of $f$ and $\lambda^t$ is some suitably chosen step size (e.g. constant or optimised via line search). This requires a representation of $f$ which is differentiable in $p$ and flexible enough to allow for a variety of shapes, including skewness and multi-modality. We can obtain such a representation by approximating $f$ as a polynomial $\hat{f}$ of some specified degree $d$, fitted to a suitable set of samples $(p^{(l)},f(p^{(l)}))$. For example, one could use a uniform grid over the parameter space that includes the boundary points. Algorithm~\ref{alg:aga} provides a summary of this method.

We note that operations such as fitting and differentiation of multivariate polynomials can be costly, even in the approximate case \cite{f2012}, whereas univariate polynomials can be processed very efficiently. To alleviate this, one may partition parameters $p_1,...,p_n$ into clusters $C_1,C_2,...$ according to their degree of correlation in $f$ (so that parameters from different clusters are independent or only weakly correlated; cf. \cite{ar2016jair}) and use separate polynomials for each cluster. If the resulting clusters are small, this can significantly reduce computational costs \cite{mw2001,bk1998}. However, care must be taken not to break important correlations between parameters, which may degrade the accuracy of parameter estimates.

\begin{algorithm}[t]
	\textbf{Given:} parameter estimate $p^{t-1}$, degree $d$ \\[-11pt]
	\begin{algorithmic}[1]
		\algsetup{linenodelimiter=:\ }
		\setstretch{1.3}
		\STATE Collect samples $D = (p^{(l)},f(p^{(l)}))$ \com{e.g. uniform grid}
		\STATE Fit polynomial $\hat{f}$ of degree $d$ to $D$
		\STATE Compute gradient $\nabla \hat{f}(p^{t-1})$ and step size $\lambda^t$
		\STATE Update estimate $p^t = p^{t-1} + \lambda^t \,\nabla \hat{f}(p^{t-1})$
	\end{algorithmic}
	\caption{Approximate Gradient Ascent}
	\label{alg:aga}
\end{algorithm}

\begin{algorithm}[t]
	\textbf{Given:} $P(p | H_i^{t-1}, \typ_j)$, represented as polynomial of deg. $d$ \\[-11pt]
	\begin{algorithmic}[1]
		\algsetup{linenodelimiter=:\ }
		\setstretch{1.3}
		\STATE Fit $\hat{f}$ to $f$ as in Algorithm~\ref{alg:aga}
		\STATE Compute polynomial product $\hat{g} = \hat{f} \cdot P(p | H_i^{t-1}, \typ_j)$
		\STATE Collect samples $D = (p^{(l)},\hat{g}(p^{(l)}))$ \com{e.g. uniform grid}
		\STATE Fit new polynomial $\hat{h}$ of degree $d$ to $D$
		\STATE Compute integral $I = \int_{p^{\min}}^{p^{\max}} \hat{h}(p) \, dp$
		\STATE Set new belief $P(p | H_i^t, \typ_j) = \hat{h} / I$
		\STATE Extract estimate $p^t$ from $P(p | H_i^t, \typ_j)$ \com{e.g. sample}
	\end{algorithmic}
	\caption{Approximate Bayesian Updating}
	\label{alg:abu}
\end{algorithm}

\begin{figure*}[t]
	\vspace{-10pt}
	\center
	\subfloat[Prior belief]{\label{fig:polybayes-a}\includegraphics[height=0.155\textheight]{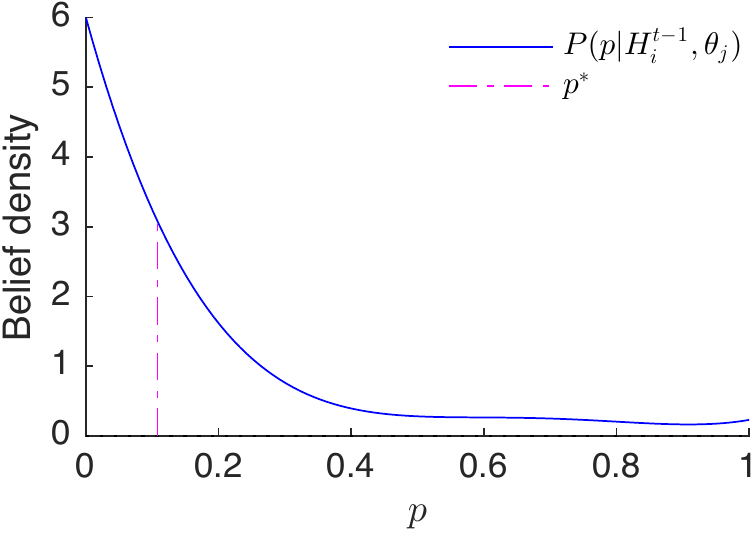}}\hspace{1.9em}
	\subfloat[$\hat{f}$ (likelihood)]{\label{fig:polybayes-b}\includegraphics[height=0.155\textheight]{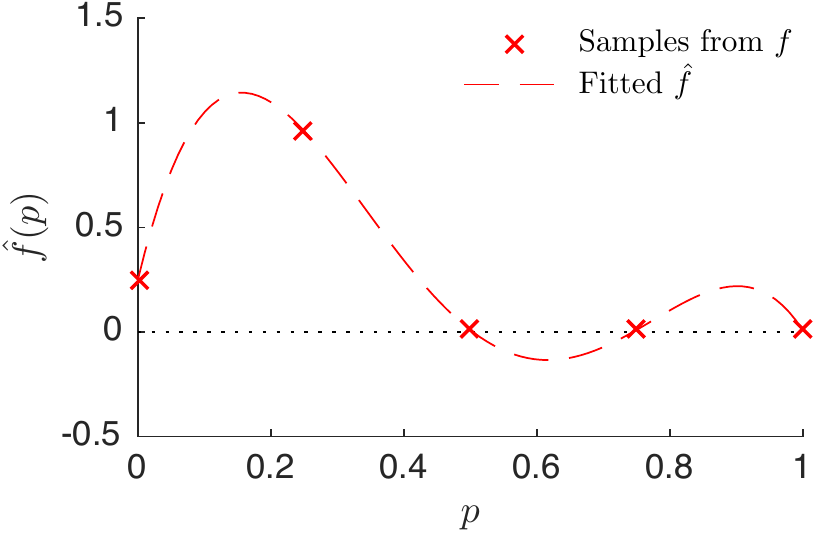}}\hspace{2.5em}
	\subfloat[Posterior belief]{\label{fig:polybayes-c}\includegraphics[height=0.155\textheight]{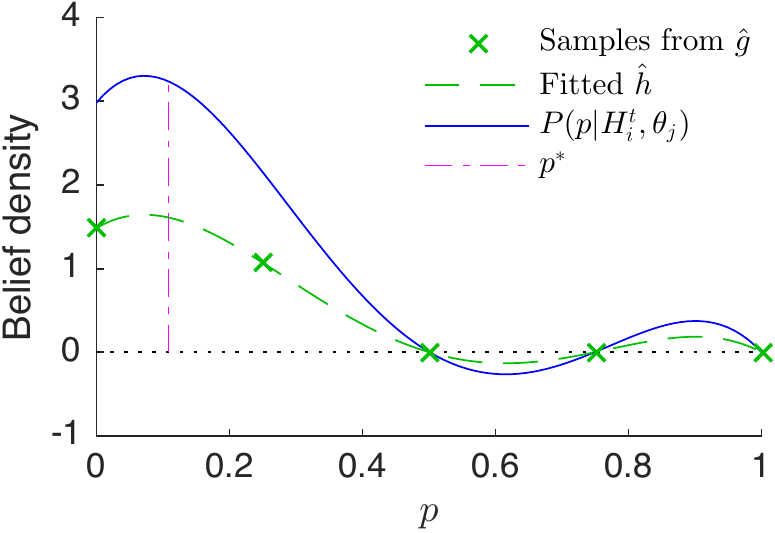}}
	\caption{Approximate Bayesian Updating for a single parameter $p \in [0,1]$ with true value $p^* = 0.11$. The polynomials have degree 4 and are fitted using 5 uniformly spaced points from the parameter space.}
	\label{fig:polybayes}
\end{figure*}

		\subsubsection{Approximate Bayesian Updating} \label{sec:estim-abu}

Rather than using $\hat{f}$ to perform gradient-based updates, we can use $\hat{f}$ to perform Bayesian updates that retain information from past updates. In addition to the belief $P(\typ_j | H_i^t)$, agent $i$ now also has a belief $P(p | H_i^t, \typ_j)$ to quantify the relative likelihood of parameter values $p$ for $\typ_j$. This new belief is represented as a polynomial of the same degree $d$ as $\hat{f}$. The Bayesian update is then constructed as follows:

After fitting $\hat{f}$, we take the convolution (i.e. polynomial product) of $P(p | H_i^{t-1}, \typ_j)$ and $\hat{f}$, resulting in a polynomial $\hat{g}$ of degree greater than $d$. To restore the original representation, we fit a new polynomial $\hat{h}$ of degree $d$ to any suitably chosen set of sample points from the convolution $\hat{g}$. Again, we could use a uniform discretisation of the parameter space. Finally, we compute the integral of $\hat{h}$ under the parameter space and divide the coefficients of $\hat{h}$ by the integral, to obtain the new belief $P(p | H_i^t, \typ_j)$. This new belief can then be used to obtain a parameter estimate, e.g. by finding the maximum of the polynomial or by sampling from the polynomial. Algorithm~\ref{alg:abu} provides a summary of this process and Figure~\ref{fig:polybayes} gives a graphical example.

While the use of polynomials allows for great flexibility, it does not come without limitations: Polynomials suffer from known instability issues in extrapolation and interpolation. Extrapolation is not of concern here since we are confined to bounded parameter spaces. However, instability of interpolation can lead to negative values between fitted samples (cf. Figure~\ref{fig:polybayes-b}). While this poses no difficulty for the calculation of maxima and sampling, it does mean that the integral in the normalisation of $\hat{h}$ has to be ``absolute'', in that any area below the zero axis is assigned a positive sign. Moreover, due to the nature of approximate fitting and finite machine accuracy, care should be taken that the samples taken from $\hat{g}$ to construct $\hat{h}$ (cf. Figure~\ref{fig:polybayes-c}) are not negative in $\hat{g}$, as otherwise negative minima may be propagated across updates, which can lead to further instabilities.

		\subsubsection{Exact Global Optimisation} \label{sec:estim-ego}

The previous methods rely on an approximation $\hat{f}$ of $f$ to perform successive updates. An alternative approach is to reason directly with $f$. In addition to avoiding the potential inaccuracies caused by the approximations, this would allow for the detection of possible discontinuities in $f(p)$ which cannot be represented by continuous polynomials.

Specifically, the estimation of parameter values can be viewed as a global optimisation problem \cite{hpt2000} in which the goal is to find a parameter setting $p^t$ with maximum probability over the history of observations $H_i^t$. Formally, the optimisation problem is defined as follows:
\begin{eqnarray} \label{eq:optim}
	\text{Find } p^t \in \arg\max_p F(p) = \prod_{\tau=1}^t P(a_j^{\tau-1} | H_i^{\tau-1}, \typ_j, p) \\
	\text{s.t. } \forall_k \ p_k \in [p_k^{\min},p_k^{\max}] \nonumber
\end{eqnarray}

Since the evaluation of the objective function $F$ for a given parameter setting $p$ can be relatively costly, one would ideally solve this problem using an optimisation method that seeks to minimise the number of evaluations. Bayesian optimisation was specifically designed for such settings and has been shown to be effective for low-dimensional problems \cite{m2012}. The idea is to use a Gaussian process \cite{rw2006} to represent uncertainty over the values of $F$. Each iteration of the method selects a new point $p$ to evaluate, according to some tradeoff criterion for exploitation (choosing points which are expected to have high values) and exploration (minimising uncertainty). A~crucial choice in this method is the form of the covariance function, which is used to measure similarity of points \cite{sla2012}.


	\section{Experimental Evaluation} \label{sec:eval}

We provide a detailed experimental evaluation of our methods in the level-based foraging domain \cite{ar2013}, which was introduced as a test domain for ad hoc teamwork \cite{skkr2010}.

		\subsection{Domain Description} \label{sec:eval-domain}

\begin{figure}[t]
	\centering
	\includegraphics[height=0.17\textheight]{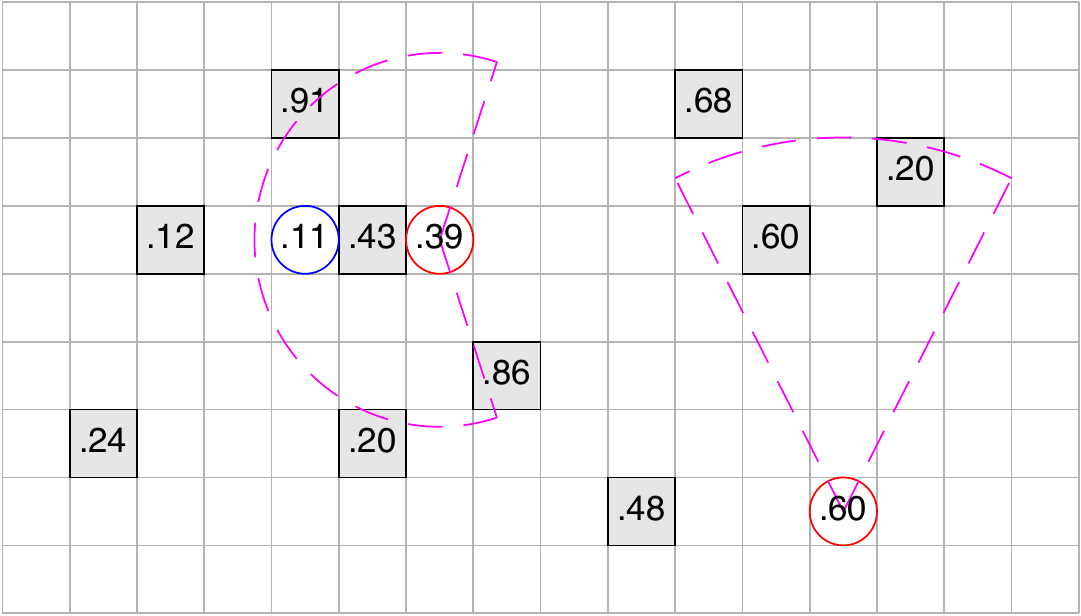}
	\caption{Level-based foraging domain. Agents are marked by circles (blue is our agent) and items are marked by grey squares. Skill levels are shown inside agents and items. The dashed magenta lines show the other agents' view cones.}
	\label{fig:foraging}
\end{figure}

The domain consists of a rectangular grid in which a team of agents must collaborate to collect a number of items in minimal time. The agents' ability to collect items is limited by \emph{skill levels}: each agent and item has an individual level which is represented by a number in the range $[0,1]$. A group of agents can collect an item if (i) they are located next to the item, (ii) they simultaneously choose the {\it load} action, and (iii) the sum of the agents' levels is at least as high as the item's level. Thus, in Figure~\ref{fig:foraging}, the two agents in the left half can jointly collect an item which individually they cannot collect. When an item is collected, it is removed from the grid and the team receives a reward of 1; in all other cases, the reward is 0 (timing will become relevant via a discount factor). In addition to the {\it load} action, each agent has 4 actions {\it N, E, S, W}, which move the agent into the corresponding direction if the target cell is empty and inside the grid. Ties are resolved by executing actions in random order.

To enforce collaboration and keep this solvable, skill levels are chosen such that all agents have levels below the highest item level, and no item has a level greater than the sum of all agent levels. Furthermore, items are placed such that the Euclidean distance between each item is greater than 1, and no item is placed at any border of the grid.

We extend this domain by adding view cones for agents, which are parameterised by a radius and angle. An agent's view cone determines which items and other agents it can see, as well as the certainty with which they are seen. The latter is calculated as the percentage (measured in $[0,1]$) with which the view cone overlaps with the grid cell occupied by an agent or item. Thus, the agent in the right half of Figure~\ref{fig:foraging} can see two items, one with certainty 1 and another one with certainty $\approx 0.85$. We assume that our agent can see the entire grid (cf. Section~\ref{sec:prel}), hence it has no view cone.

		\subsection{Hypothetical Type Space} \label{sec:eval-types}

The hypothetical type space $\Typ_j$ consists of four types which are all based on the template given in Algorithm~\ref{alg:template}. The template uses three parameters: $p_1 \in [0,1]$ specifies the agent's skill level; $p_2 \in [.1,1]$ specifies the agent's view radius as $p_2 \sqrt{w^2+h^2}$, where $w$ and $h$ are the width and height of the grid; and $p_3 \in [.1,1]$ specifies the view angle as $p_3 2\pi$. The parameters $p_2, p_3$ are used in the \textsc{VisibleAgentsAndItems} routine, which returns two sets containing the visible agents and items with a view certainty of 0.1 or higher. The parameter $p_1$ is used in the \textsc{ChooseTarget} routine, which returns a specific target out of the visible agents and items.

The four types in $\Typ_j$ differ from each other in their specification of the \textsc{ChooseTarget} routine:
\begin{itemize}[leftmargin=10pt,itemsep=0pt]
	\item $\typ_j^{L1}$: if items visible, return furthest\footnote{\small We found that choosing the furthest item/agent penalises wrong parameter estimates more than choosing closest ones, since the latter is invariant to overestimation of view cone parameters.} one; else, return $\emptyset$
	\item $\typ_j^{L2}$: if items visible, return item with highest level below own level, or item with highest level if none are below own level; else, return $\emptyset$
	\item $\typ_j^{F1}$: if agents visible but no items visible, return furthest agent; if agents and items visible, return item that furthest agent would choose if it had type $\typ_j^{L1}$; else, return $\emptyset$
	\item $\typ_j^{F2}$: if agents visible but no items visible, return agent with highest level above own level, or furthest agent if none are above own level; if agents and items visible, select agent as before and return item that this agent would choose if it had type $\typ_j^{L2}$; else, return $\emptyset$
\end{itemize}

Intuitively, types $\typ_j^{L1}$ and $\typ_j^{L2}$ can be viewed as \emph{leaders}: they choose targets on their own and expect others to follow their lead. Conversely, types $\typ_j^{F1}$ and $\typ_j^{F2}$ can be viewed as \emph{followers}: they assume other agents know best and attempt to follow their lead. The leader and follower types are further distinguished by whether they consider skill levels.

The internal state of the template is defined by a memory $\Mem$ for the current destination (x/y position) which the agent is trying to reach. Once the destination in $\Mem$ has been reached, the template chooses a new destination using the \textsc{ChooseTarget} routine. Thus, the contents of $\Mem$ is directly affected by the parameters, and we can classify them as non-Markovian (cf. Section~\ref{sec:markov}).

Finally, Line~\ref{alg:mix} in Algorithm~\ref{alg:template} is a simple way of guaranteeing the assumption that the set of actions with positive probability is invariant of the parameter values (see last paragraph before Section~\ref{sec:select}).


\begin{algorithm}[t]
	\vspace{1pt}
	\textbf{Parameters:} skill level $p_1$, view radius $p_2$, view angle $p_3$ \\[1pt]
	\textbf{Initialise:} destination memory $\Mem \gets \emptyset$ \\[1pt]
	\textbf{Repeat} for each $t$:
	\begin{algorithmic}[1]
		\algsetup{linenodelimiter=:\ }
		\setstretch{1.1}
		\STATE \com{Select destination}
		\STATE $\Loc \gets \text{own location}$,\, $\Des \gets \emptyset$
		\IF{$\Mem \neq \emptyset$ and $\Loc \neq \Mem$}
			\STATE $\Des \gets \Mem$ 
		\ELSE
			\STATE $(A,I) \gets \textsc{VisibleAgentsAndItems}(\Loc)$
			\STATE $\Tar \gets \textsc{ChooseTarget}(A,I)$
			\IF{$\Tar \neq \emptyset$}
				\STATE $\Des \gets \Tar$
			\ENDIF
		\ENDIF
		\STATE Save destination in memory: $\Mem \gets \Des$
		\vspace{3pt}
		\STATE \com{Assign action probabilities}
		\IF{$\Des = \emptyset$}
			\STATE Assign probability 0.25 to each move action
		\ELSE
			\IF{$\Des$ is item and $\Loc$ is next to $\Des$}
				\STATE Assign probability 1 to {\it load} action
			\ELSE
				\STATE Use $A^*$ \cite{hnr1968} to find path from $\Loc$ to $\Des$
				\STATE Assign probability 1 to first move action in path
			\ENDIF
		\ENDIF
		\STATE Add probability 0.01 to each action and normalise \label{alg:mix}
	\end{algorithmic}
	\caption{Template for foraging types}
	\label{alg:template}
\end{algorithm}

		\subsection{Experimental Setup}

We tested various configurations of~Algorithm \ref{alg:main}. For the selection of types for parameter updates ($\Up$), we tested updating all types in $\Typ_j$, sampling a single type from $\Typ_j$ using the belief $P(\typ_j | H_i^{t-1})$ (Section~\ref{sec:select-post}), and sampling a single type from $\Typ_j$ using a bandit algorithm (Section~\ref{sec:select-bandit}). A number of bandit algorithms were tried in preliminary experiments, including UCB1 \cite{acf2002}, EEE \cite{fm2004}, S \cite{kmrv1998}, Exp3 \cite{acfs1995}, and Thompson sampling \cite{r1933}. All reported results are based on UCB1, which achieved the best performance.

For the estimation of parameter values, we tested Approximate Gradient Ascent (AGA), Approximate Bayesian Updating (ABU), and Exact Global Optimisation (EGO). AGA and ABU used univariate polynomials of degree 4 for each parameter, which were fitted using 5 uniformly spaced points over the parameter space (as shown in Figure~\ref{fig:polybayes}). AGA optimised the step size $\lambda^t$ in each update using backtracking line search (with the search parameters set to 0.5/0.5). ABU used uniform initial beliefs $P(p | H_i^0, \typ_j)$ for each type $\typ_j \in \Typ_j$ and generated parameter estimates by averaging over 10 samples taken from $P(p | H_i^t, \typ_j)$ (which we found to be more robust than taking the maximum). EGO was implemented using Bayesian optimisation with the ``expected improvement'' search criterion \cite{m2012} and squared exponential covariance with automatic relevance detection \cite{rw2006}. The number of points evaluated by EGO (cf. \eqref{eq:optim}) was limited to 10.

All configurations used uniform initial beliefs $P(\typ_j | H_i^0)$ over the set $\Typ_j$ (specified in Section~\ref{sec:eval-types}) and random initial parameter estimates for each $\typ_j \in \Typ_j$. In each time step, Monte Carlo Tree Search (MCTS), specifically UCT \cite{ks2006}, was used to compute optimal actions with respect to the beliefs and types. Each rollout in the tree search used the current belief $P(\typ_j | H_i^t)$ to sample a type $\typ_j \in \Typ_j$ which was used for the entire rollout. Each time step generated 300/500 rollouts in the 10x10/15x15 worlds, respectively (see below), which we found to be robust numbers. Each rollout was over a horizon of 100 time steps, and the rewards accumulated during a rollout were discounted with a factor of 0.95. Subtrees from previous time steps were reused to accelerate the tree search.

The configurations were tested in two different sizes of the level-based foraging domain: a 10x10 world with 2 agents and 5 items, and a 15x15 world with 3 agents and 10 items (so our agent reasons about the types and parameters of two other agents). Each configuration was tested in the same sequence of 500 instances, which were generated as follows: First, we set random initial positions and skill levels for each agent and item, subject to the constraints noted in Section~\ref{sec:eval-domain}. Then, for each agent not under our control, we randomly selected its true type $\typ_j^*$ from the type space $\Typ_j$ and completed its parameter setting by choosing random values for the view cone parameters. Finally, for each $\typ_j \in \Typ_j$, we sampled random initial parameter estimates which were used by the tested configuration. Instances of the 10x10/15x15 world were run for a maximum of 100/150 time steps, respectively.

We used two baselines to facilitate the comparison of our methods: \rnd, which used fixed random parameter values for each type, and \cor, which used the correct parameter values for the true type and fixed random parameter values for all other types (baselines did not update parameters).

		\subsection{Results}

\begin{figure}[t]
	\centering
	\subfloat[10x10 world, 2 agents, 5 items]{\includegraphics[height=0.129\textheight]{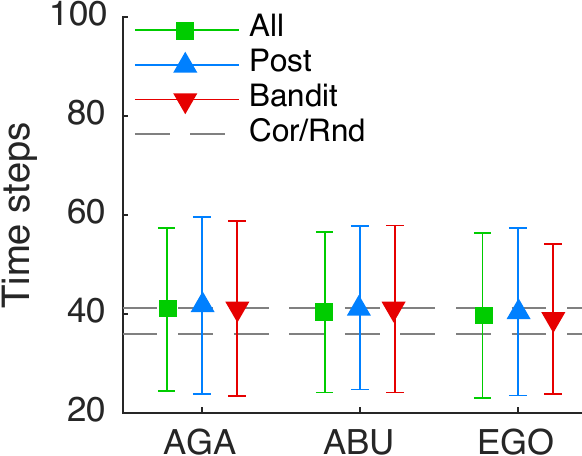}\hspace{10pt}\includegraphics[height=0.129\textheight]{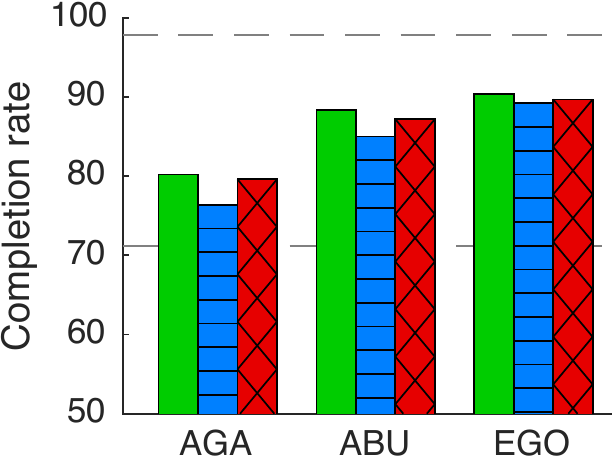}}\\
	\subfloat[15x15 world, 3 agents, 10 items]{\includegraphics[height=0.129\textheight]{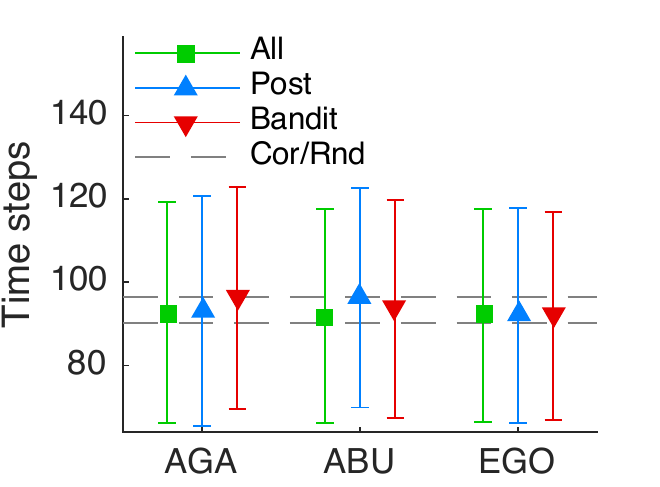}\hspace{10pt}\includegraphics[height=0.129\textheight]{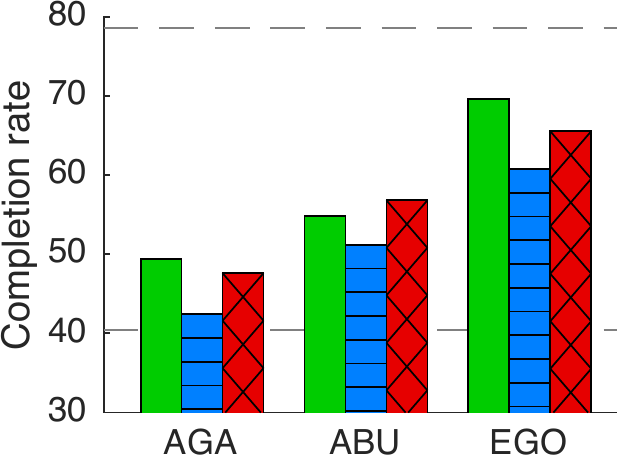}}
	\caption{Time steps required in completed instances (means and standard deviations) and completion rates for the tested methods. Results are averaged over 500 instances in each world. Dashed lines mark the baseline performances, where \cor\ had lowest time steps and highest completion rates.}
	\label{fig:results}
\end{figure}

Figure~\ref{fig:results} shows the average number of time steps and the completion rates for each of the tested configurations and world sizes. The completion rate is the percentage of instances which were completed successfully (i.e. all items collected) within the given amount of time. The average time steps are for completed instances. To put the results into perspective, we will begin by discussing the results of the two baselines, \cor\ and \rnd. (In the following, all significance statements are based on paired t-tests with a 5\% significance threshold.)

\begin{textblock}{10}(55,10.5)
	\includegraphics[height=0.047\textheight]{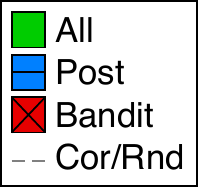}
\end{textblock}
\begin{textblock}{10}(49.2,18.8)
	\includegraphics[height=0.047\textheight]{completion_legend.pdf}
\end{textblock}

\vspace{2.5pt}

The first observation is that there was only a small difference between \cor\ and \rnd\ in their average number of time steps for completed instances, with margins of less than 10 time steps in both world sizes. This may seem surprising, given that the random parameter settings used by \rnd\ can lead to significantly different predictions than the correct settings. However, in instances which were completed by both baselines, we found that the MCTS planner was robust enough to ``absorb'' the differences, in that it often produced similar courses of actions despite the differences. On the other hand, there were substantial differences in the completion rates of \cor\ and \rnd, dropping from 98\% to 71\% in the 10x10 world and 79\% to 41\% in the 15x15 world, respectively. We found that the random parameter settings used by \rnd\ often led to predictions that fooled \rnd\ into taking the wrong actions without ever realising it, thus inducing an infinite cycle which the agent never escaped. This effect has been described previously as ``critical type spaces'' \cite{ar2014}. Given the means and standard deviations of time steps shown in Figure~\ref{fig:results}, one can see that simply increasing the maximum allowed time steps per instance would not significantly affect \rnd's ability to complete instances.

We now turn to a comparison of our proposed methods. Most notably, the results show that updating a single type in each time step achieved comparable performance to updating \emph{all} types in each time step, albeit at only a fraction (approximately $\frac{1}{4}$th, since $|\Typ_j| = 4$) of the computation time. Moreover, bandit selection significantly outperformed posterior selection in all tested configurations, except for EGO in the 10x10 world, where the two were equivalent. We found that this difference was due to the fact that posterior selection tended to exploit more greedily than bandit selection, because the beliefs $P(\typ_j | H_i^t)$ often placed high probability on certain types early on in the interaction. In contrast, bandit selection was more exploratory because the rewards defined in Section~\ref{sec:select-bandit} tended to be more uniform across types than beliefs. Given that the distributions underlying these rewards were not stationary, it is worth pointing out that bandit algorithms which were specifically designed for changing distributions (e.g. \cite{fm2004,acfs1995}) did not perform better than those which assume stationary reward distributions.\footnote{\small The analysis in \cite{ks2006} provides some insights into the performance of UCB1 for non-stationary (``drifting'') reward distributions.} These results show that our approach of viewing the selection of types as a decision problem, balancing exploitation and exploration, can be effective in practice.

\begin{figure}[t]
	\centering
	\includegraphics[height=0.115\textheight]{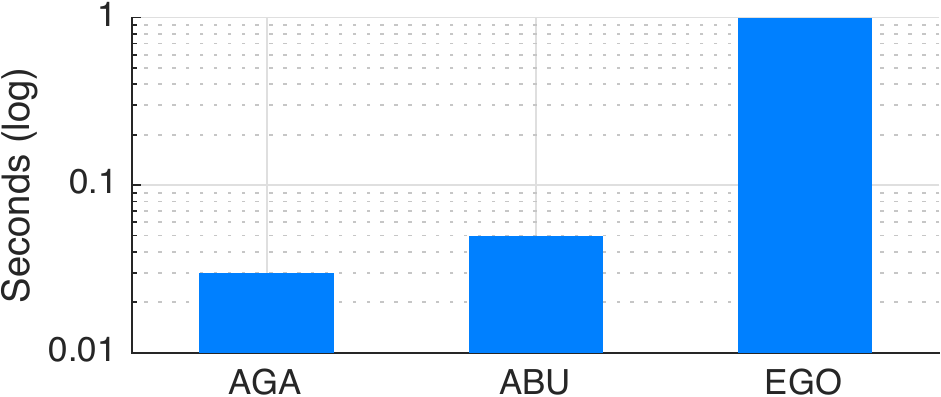}
	\caption{Average seconds (log-scale) needed per parameter update for one type. Measured in Matlab R2015b on a UNIX dual-core machine with 2.66 GHz per core.}
	\label{fig:time-secs}
\end{figure}

Regarding the different estimation methods, the results show a gradual improvement from AGA to ABU to EGO. AGA performed worst because the gradient update used in AGA did not retain information from past updates. Thus, its estimates were dominated by the most recent observations, which often prevented convergence to good estimates. In addition, AGA and ABU's ability to estimate parameters was hindered by the fact that they used individual polynomials for the parameters, thus ignoring possible parameter correlations at the benefit of reduced computation time. EGO, due to its ability to detect parameter correlations and discontinuities, achieved the best performance in our experiments. We note that the results shown for EGO are for a maximum of 10 evaluated points. We were able to drive its performance up by increasing the number of evaluated points, approaching the performance of the \cor\ baseline in both worlds. However, this performance came at a significant cost in computation time (cf. Figure~\ref{fig:time-secs}): while AGA and ABU needed on average about 0.03 and 0.05 seconds per update, EGO needed about 1 (2.3) seconds per update when evaluating 10 (20) points, which increased slowly for longer histories. Thus, ABU provided the best tradeoff between task completion and computation time. However, the time requirements of EGO may be reduced drastically by using a more efficient implementation of Bayesian optimisation, e.g. \cite{bayesopt2014}.

\begin{figure}[t]
	\centering
	\subfloat[Parameter $p_1$ (skill level)]{\includegraphics[height=0.136\textheight]{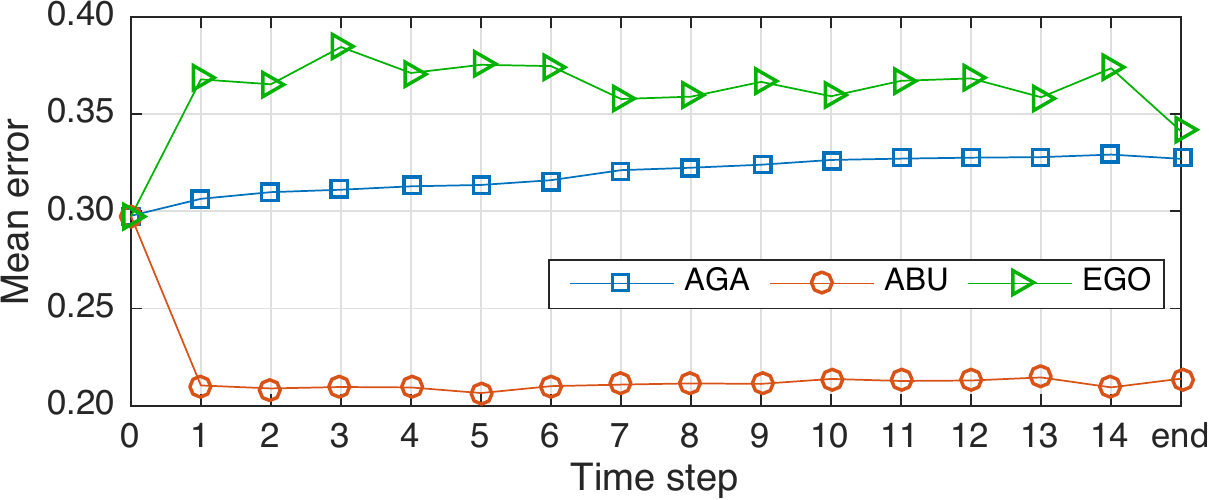}}\\
	\subfloat[Parameter $p_2$ (view radius)]{\includegraphics[height=0.136\textheight]{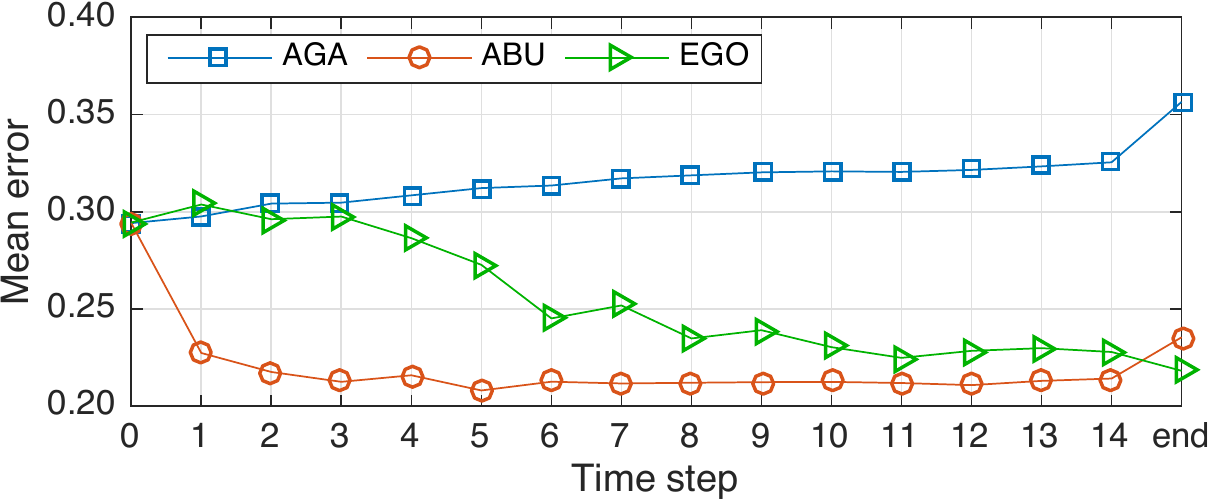}}\\
	\subfloat[Parameter $p_3$ (view angle)]{\includegraphics[height=0.136\textheight]{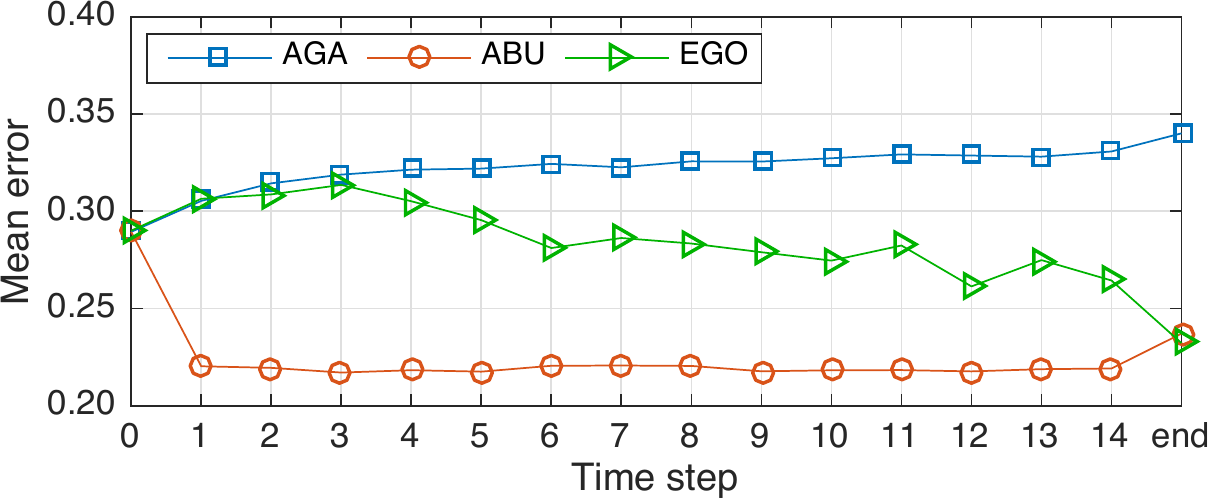}}
	\caption{Mean error in parameter estimates for the true type $\typ_j^*$ in the 15x15 world (updating all types in each time step), averaged over 500 instances and both other agents. The error at time $t$ is defined as the absolute difference $|p_k^* - p_k^t|$. Errors are shown for the first 15 and last time steps of an instance.}
	\label{fig:error}
\end{figure}

Figure~\ref{fig:error} shows the mean error in the parameter estimates for the true type $\typ_j^*$. The figure shows that AGA's estimation errors increased slowly over time. One reason for this was that $f$ (i.e. the action probabilities of types with respect to parameters; cf. Section~\ref{sec:estim}) was often multi-modal and hence non-convex, causing the gradient to point away from the true parameter values. Another reason was that $f$ could change drastically between time steps, which in some cases had a similar ``trapping'' effect on the gradient. Nonetheless, AGA still managed to produce good estimates in some of the instances. A different picture is shown for ABU: its mean errors dropped substantially after the first time step and remained stable after. This shows that ABU was able to effectively retain information from past updates, through its conjugate polynomial update. While EGO did also retain information from past observations, its estimates were less stable than ABU's estimates, often jumping radically between different values. This was a result of the search strategy used in Bayesian optimisation and the fact that it only evaluated 10 points in each update, which can cause it to find different solutions after each new observation. An interesting observation is that EGO seemed to differentiate between parameters, with substantially different mean errors for the individual parameters. This, too, was a result of its search strategy, which can concentrate on certain parameters if they lead to better solutions. Thus, $p_1$ (the skill level) seemed to be less relevant than $p_2/p_3$ (the view cone parameters). Given that ABU's mean error was substantially lower than EGO's mean error, it may be surprising that EGO still outperformed ABU in completion rates. However, a closer inspection showed that EGO more often estimated the right \emph{combination} of parameter values (i.e. it recognised correlations in parameters), which in many cases was crucial for the correct planning of actions.

\begin{figure}[t]
	\centering
	\includegraphics[height=0.15\textheight]{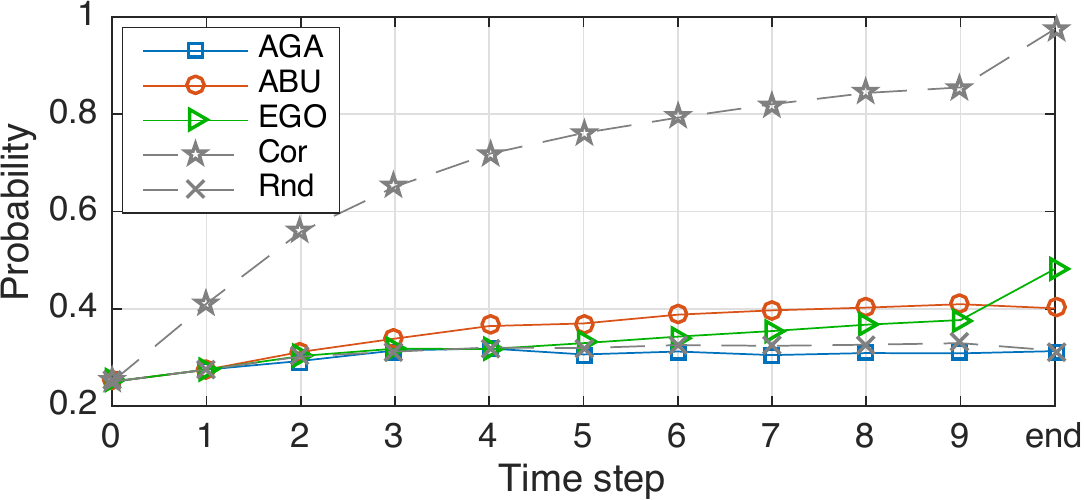}
	\caption{Average belief $P(\typ_j^* | H_i^t)$ for the true type $\typ_j^*$ in the 10x10 world (updating all types in each time step). Probabilities are averaged over 500 instances and shown for the first 10 and last time steps of an instance.}
	\label{fig:post}
\end{figure}

Finally, Figure~\ref{fig:post} shows the evolution of beliefs in the 10x10 world (the same picture was obtained in the 15x15 world). The correct baseline \cor\ had a robust convergence to the true type, with an average final probability of 0.975 for the true type. In contrast, the random baseline \rnd\ converged in many cases to an incorrect type, with an average final probability of 0.314 for the true type. The corresponding probabilities produced by our methods were 0.313 for AGA, 0.401 for ABU, and 0.482 (0.574) for EGO with 10 (50) evaluated points. Thus, AGA did not improve belief convergence over \rnd\ while ABU and EGO produced statistically significant improvements, albeit still a long way from \cor. By the end of an instance, all methods placed most of their belief mass on one type, with average maximum probabilities (over any type) in the 0.9x range. These numbers show that parameter estimates that deviate from the true values can have a significant impact on the evolution of beliefs. As our data show, convergence to the true type correlates with (and causes) higher completion rates.


	\section{Discussion}

		\subsection{A Note on Belief Merging}

A central feature of keeping beliefs over a set of behaviours is a property called \emph{belief merging} \cite{kl1993}. Under a condition of ``absolute continuity'', this property entails that the believed distribution over future play converges in a strong sense to the true distribution induced by the true behaviour. One may ask if this property also holds in our method, given that \eqref{eq:bayesup} may use different parameter estimates in each update.

The simple answer to this question is no, because changing the parameter estimates means that the beliefs effectively refer to a different type space in the original result \cite{kl1993}. Would a method that uses distributions over parameter values rather than point estimates inherit the belief merging property? It can be shown that the answer here, too, is negative, and we provide an example below (we assume basic familiarity with the work of Kalai and Lehrer \cite{kl1993}):

Suppose agent $j$ can choose between two actions. Its true type, $\typ_j^*$, is to choose action~1 with probability $\delta$ and action~2 with probability $1-\delta$. Assume that agent $i$ knows $\typ_j^*$ but not the value of the $\delta$ parameter, and so maintains a continuous distribution over the interval $[0,1]$. The probability measures $\mu$ and $\mu'$ over play paths are induced in the usual way \cite{kl1993} from the true type and the distribution, respectively. Now, consider the set $\Omega$ consisting of all infinite play paths in which action 1 has limit frequency $\delta$. We have $\mu(\Omega) = 1$, since $\typ_j^*$ can only realise paths in $\Omega$, but $\mu'(\Omega) = 0$ due to the diffused distribution over $\delta$. Thus, the absolute continuity condition is violated and belief merging does not materialise (absolute continuity is in fact necessary for belief merging \cite{kl1994}).

Nonetheless, it has been argued that absolute continuity and the resulting convergence (which implies accurate prediction of \emph{infinite} play paths \cite{kl1993}) are too strong for practical applications \cite{dg2006,n2005,kl1994}. It is easy to see that the ABU and EGO methods described in Section~\ref{sec:estim} would converge point-wise to the correct parameter value in the above example.

		\subsection{Related Work} \label{sec:relwork}

Several works proposed methods which maintain Bayesian beliefs over a set of possible behaviours or types \cite{acr2016aij,bsk2011,gd2005,sbl2005,cb2003,cm1999}. Some methods assume discrete (usually finite) type spaces \cite{ar2013,bsk2011,cm1999} while others assume continuous type spaces \cite{sbl2005,cb2003}. Our work can be viewed as bridging these methods by doing both: we maintain beliefs over a finite set of types, and we allow each type to have continuous parameters. Moreover, our methods can deal with any parameterisation, while the methods proposed in \cite{sbl2005,cb2003} are specific to parameters of the used distributions (e.g. Dirichlet).

Classical methods for opponent modelling assume a fixed model structure (e.g. a decision tree or finite-state machine) and attempt to fit the model parameters based on observed actions (e.g. \cite{bskr2013,lasb2004,cm1996}). Because such models may involve many parameters, the learning process may need many observations to produce useful fits. This is in contrast to type-based methods, in which types are blackbox functions and we only ``fit'' one probability for each type. The latter can lead to rapid adaptation, but may not be as flexible as classical methods. Here, too, our work can be viewed as a hybrid between the two approaches: in addition to fitting probabilities over types we now also fit parameters within types, giving them greater flexibility, but the number of such parameters is usually lower than that found in classical methods.

Our proposed method is in part inspired by methods of selective inference in dynamic Bayesian networks \cite{ar2016jair}. In our work, we selectively choose types whose parameter values we wish to infer. However, the selection of types is viewed as a decision problem whereas the selective inference in \cite{ar2016jair} is predetermined by the structure of the network.


		\subsection{Conclusion \& Outlook} \label{sec:conc}

This work extends the type-based interaction method by allowing an agent to reason about both the relative likelihood of types \emph{and} the values of any bounded continuous parameters within types. A key element in our approach to minimise computation costs is to perform \emph{selective updates} of the types' parameter estimates after new observations are made. Moreover, our proposed methods for the estimation of parameter settings can be applied to any continuous parameters in types, without requiring additional structure in type specifications. We evaluated our methods in detailed experiments, showing that they achieved substantial improvements in task completion rates compared to random estimates, while updating only a single parameter estimate in each time step.

There are several potential directions for future research. Our experiments showed that parameter estimates can have a significant effect on the evolution of beliefs over types. However, we do not currently have a formal theory that characterises the interaction between parameter estimates and beliefs. Such a theory might have useful implications for the selection of types and the derivation of estimates. Furthermore, our methods assume that we can observe (or derive) the chosen actions and observations of other agents. A useful generalisation of our work would be to also account for possible uncertainties in such observations, e.g. \cite{pg2017}. Finally, further enhancements of our methods could be made. For instance, another approach to select types for updates might be to estimate the impact that updating a particular type may have on our beliefs and future actions. However, such methods can be computationally expensive, even in the myopic approximate case \cite{cb2003}.

	\small

\vspace{1em}

\noindent\textbf{Acknowledgements:} This work took place in the Learning Agents Research Group (LARG) at UT Austin. LARG research is supported in part by NSF (CNS-1330072, CNS-1305287, IIS-1637736, IIS-1651089), ONR (21C184-01), AFOSR (FA9550-14-1-0087), Raytheon, Toyota, AT\&T, and Lockheed Martin. Peter Stone serves on the Board of Directors of, Cogitai, Inc. The terms of this arrangement have been reviewed and approved by The University of Texas at Austin in accordance with its policy on objectivity in research. Stefano Albrecht is supported by a Feodor Lynen Research Fellowship from the Alexander von Humboldt Foundation.

	\bibliographystyle{abbrv}
	\bibliography{aamas17}

\end{document}